\documentclass[pdflatex,sn-mathphys-num]{sn-jnl}

\usepackage{graphicx}%
\usepackage{subcaption} 
\usepackage{multirow}%
\usepackage{amsmath,amssymb,amsfonts}%
\usepackage{amsthm}%
\usepackage{mathrsfs}%
\usepackage[title]{appendix}%
\usepackage{xcolor}%
\usepackage{textcomp}%
\usepackage{manyfoot}%
\usepackage{booktabs}%
\usepackage{algorithm}%
\usepackage{algorithmicx}%
\usepackage{algpseudocode}%
\usepackage{listings}%
\usepackage{url}
\usepackage{lineno}

\theoremstyle{thmstyleone}%

\theoremstyle{thmstyletwo}%

\theoremstyle{thmstylethree}%
\newtheorem{definition}{Definition}%

\begin{document}


\title[]{Temporal Dynamics of Emotions in Italian Online Soccer Fandoms}





\author*[1]{\fnm{Salvatore} \sur{Citraro}}\email{salvatore.citraro@isti.cnr.it}

\author[2]{\fnm{Giovanni} \sur{Mauro}}\email{giovanni.mauro@sns.it}

\author[3]{\fnm{Emanuele} \sur{Ferragina}}\email{emanuele.ferragina@sciencespo.fr}

\affil*[1]{\orgname{CNR-ISTI}, \orgaddress{\street{Via Giuseppe Moruzzi, 1}, \city{Pisa}, \country{Italy}}}

\affil[2]{\orgname{Scuola Normale Superiore}, \orgaddress{\city{Pisa}, \country{Italy}}}

\affil[3]{\orgname{SciencesPo}, \orgaddress{\street{Rue Saint Guillaume, 27}, \city{Paris}, \country{France}}}


\abstract{This study investigates the emotional dynamics of Italian soccer fandoms through computational analysis of user-generated content from official Instagram accounts of 83 teams across Serie A, Serie B, and Lega Pro during the 2023-24 season. By applying sentiment analysis to fan comments, we extract temporal emotional patterns and identify distinct clusters of fan bases with similar preseason expectations. Drawing from complex systems theory, we characterize joy as displaying anti-bursty temporal distributions, while anger is marked by pronounced bursty patterns. Our analysis reveals significant correlations between these emotional signals, preseason expectations, socioeconomic factors, and final league rankings. In particular, the burstiness metric emerges as a meaningful correlate of team performance; statistical models excluding this parameter show a decrease in the coefficient of determination of $32\%$. These findings offer novel insights into the relationship between fan emotional expression and team outcomes, suggesting potential avenues for research in sports analytics, social media dynamics, and fan engagement studies.}






\maketitle


\section{Introduction}\label{sec1}

Computational methods in social sciences have revolutionized our understanding of human behavior, offering new insights into phenomena such as the structure of social networks and large-scale collective changes \cite{girvan2002community, lazer2009computational, edelmann2020computational, tardelli2024temporal}.
An important aspect of this transformation is the capability to study sentiment and emotion at scale, allowing researchers to trace collective emotional dynamics by identifying group-based and recurring narratives \cite{hutto2014vader, zhao2016sentiment, blei2003latent, grootendorst2022bertopic}.
In this context, soccer fandoms provide a unique and rich environment to explore group-based emotions \cite{goldenberg2016process, goldenberg2020collective}.
Whilst computational methods have been successfully applied in understanding soccer from the perspective of player performance, game outcomes, injuries, scoring probability \cite{cintia2015harsh, rossi2018effective, pappalardo2019playerank, cintia2021interactive}, much less attention has been paid to the emotional experiences of fans. 
Recent studies \cite{schumaker2016predicting, pacheco2016characterization, lucas2017goaalll, thonhauser2020emotional, guzman2021towards, buongiovanni2022will, attie2023getting, wang2023making, citraro2024burstiness} have started to investigate this domain, but a significant gap remains in the application of rigorous computational methods to fully capture the nature and dynamics of emotions in online soccer fandoms.

Social media have transformed the way fans engage, reshaping traditional soccer cultures \cite{woods2022changing} by shifting communities from physical to digital spaces and, in the process, deterritorializing fandoms \cite{lawrence2018hyperdigitalization}.
Yet, it remains to be established the extent to which digital spaces are different from traditional and physical spaces.
Online platforms might help expand the reach and prestige of top teams (as shown by the growing globalisation of merchandising and the attention European top clubs receive in countries where the interest for football was not considerable in the recent past) \cite{lawrence2018hyperdigitalization}; even so, the nature of fan identity, especially within lower leagues, remains strongly tied to local and community-based features \cite{mainwaring2012we}.
Moreover, it is unclear whether online fan interactions reflect traditional emotional dynamics related to sport dynamics, which are characterized by intense post-game discussions involving enthusiasm, frustrations, opinions, \textit{what-if}s \cite{biscaia2012effects}.
This digital shift raises questions about how online interactions compare with the long-standing traditions of soccer fandoms.

In this work, we explore the emotional dynamics of digital soccer fandoms within the Italian league system, including Serie A, Serie B, and Lega Pro, during the 2023-24 season.
We aim to capture the complex patterns underlying the social behavior of Italian fans online, by answering the following research questions: \emph{(i)} Do different pools of fans exhibit similar patterns in how they emotionally support their teams?; \emph{(ii)} Can these eventual patterns, together with preseason expectations and other deeply rooted socio-economic factors, correlate with the seasonal performance of the relative teams?

By representing emotions as measurable time-dependent signals, we address question \emph{(i)} employing approaches stemming both from unsupervised machine learning and complex system analysis, namely clustering methods for time series \cite{shumway2000time} and burstiness for describing in/homogeneous temporal patterns in signal sequences \cite{goh2008burstiness}.
Segmenting users' emotional trends, our clustering analysis uncovers groups of similar fandoms sharing homogeneous features that proxy preseasonal fandoms' expectations.
Burstiness reveals regularities in expressions of joy and heterogeneous patterns of bursty/anti-bursty anger, related to the specific characteristics of each fandom.

Finally, to address question \emph{(ii)}, whether these measures influence the relationship between fandom behavior and the seasonal performance of their respective teams, we employ multiple regression analysis. This approach enables us to assess how emotional patterns, such as burstiness, correlate with final team rankings. By analyzing these dynamics, we demonstrate how emotional signals can help explain variations in team performance.

The remainder of the paper is organized as follows.
Section \ref{sec:mm} describes data collection and the methods used for clustering, burst detection, and multi-regression.
Section \ref{sec:results} focuses on the results achieved using the previously described data and methods.
Section \ref{sec:disc} concludes the paper by discussing the results.

\section{Materials and Methods}
\label{sec:mm}

\subsection{Data Collection and Pre-processing}
\label{subsec:data}

We collect user comments from the official Instagram accounts of 83 Italian teams that participated in the 2023-24 seasons of Serie A, Serie B, and Lega Pro; the top three divisions in the Italian soccer system.
Teams' accounts share content such as match updates, including squad list, half- and full-time scores, highlights, etc.
We refer to this type of content as ``posts". 
The frequency of posting activity varies between teams; cf. \ref{secA1.1}.

Users write comments in response to teams' posts and express their opinions, feelings, and reactions to the post content.
We refer to users' activities as ``comments".
Considering that some posts can contain one or a very few number of comments, we detect individual teams' thresholds to discard non-relevant posts.
In other words, we retain posts that have at least a number of comments equal to the median of the distribution of the team account, cf. \ref{secA1.1}.

We perform an emotion detection task using \textit{feel-it}\footnote{\url{https://github.com/MilaNLProc/feel-it}}, a python library tailored for the Italian language \cite{bianchi2021feel}.
Comments are classified into one of four base emotions: joy, anger, sadness, and fear.
Consequently, we count the percentage of emotions conveyed by the comments on each post.

Each post is further labeled according to the absence/non maximal (0) or maximal (1) presence of an emotion, for each emotion.
For example, a post containing 75\% of comments expressing anger, 15\% expressing joy, 5\% sadness, and 5\% fear, is labelled 0 in the dimensions of joy, sadness and fear, and 1 in the dimension of anger, thus being considered as an ``emotional event" of this latter dimension.
See Appendix \ref{secA1.2} for other information about this representation.

\noindent Formally, let $\mathcal{E}$ be the set of emotions $\mathcal{E}=\{joy, anger, sadness, fear\}$.

\begin{definition}[\textbf{Emotional Time Series Dataset}]
An emotional time series dataset $\mathcal{P}=\{P_1, P_2, ..., P_n\} \in \mathbb{R}^{n \times |\mathcal{E}| \times m}$ is a collection of $n$ emotional time series.
\end{definition}

\begin{definition}[\textbf{Emotional Time Series}]
An emotional time series $P$ is a collection of $|\mathcal{E}|$ emotional signals $P=\{p_{joy}, p_{anger}, p_{sadness}, p_{fear}\} \in \mathbb{R}^{|\mathcal{E}| \times m}$.
\end{definition}

\begin{definition}[\textbf{Emotional Signal}]
An emotional signal, like $p_{joy}$, is a sequence of $m$ real-valued observations as the percentage of comments classified as joy within a post, i.e., $p_{joy} = [p_1, p_2, ..., p_m] \in \mathbb{R}^{m}$; the same holds for the other three emotions.
\end{definition}

Moreover, for an event-based emotional representation, we consider a binary labeling of emotional events for each emotion in $\mathcal{E}$, where an emotional event is defined by the maximal presence of a specific emotion.

\begin{definition}[\textbf{Binary Emotional Signal}]
Given a post $P_i$, its emotional signal $p_e \in P_i$ for an emotion $e \in \mathcal{E}$ is transformed into a binary emotional signal $s_e = [s_1, s_2, ..., s_m] \in \{0,1\}^m$, assuming value 1 if emotion $e$ is maximal within the post, 0 otherwise.
\end{definition}

\noindent This transformation results in a dataset of event-based emotional time series.

\begin{definition}[\textbf{Event-based Emotional Time Series}]
An event-based emotional time series $S$ is a collection of $|\mathcal{E}|$ binary emotional signals $S_i = \{s_{joy}, s_{anger}, s_{sadness}\}$.
\end{definition}

\begin{definition}[\textbf{Event-based Emotional Time Series Dataset}]
An event-based emotional time series dataset $\mathcal{S}=\{S_1, S_2, ..., S_n\} \in \mathbb{R}^{n \times |\mathcal{E}| \times m}$ is a collection of n event-based emotional time series. 
\end{definition}

\noindent These time series representations guarantee the privacy of individual users \cite{monreale2010movement}.
Textual content from users can not be traced or monitored, as all comments are aggregated into an emotion distribution of the posts, and our analysis is limited to the post level.


\subsection{External Features}
\label{subsec:features}

Additionally, our analysis incorporates key contextual variables for each time series: per capita income (PCI), unemployment rate (U), welfare spending (W)\footnote{PCI, U, and W have been collected using Istat data: \url{https://www.istat.it/}}, geographical location (GEO), team heritage $R_H$\footnote{This variable is gathered following the Italian criterion for team reputation: \url{https://it.wikipedia.org/wiki/Tradizione_sportiva}}, and team market value (MV)\footnote{Since market values fluctuate throughout the season, we consider only their values at the beginning of the 2023/24 seasons --- specifically, the day before each league's kickoff, retrieving them from \url{https://www.transfermarkt.it/}}. These data enrich our analytical framework by providing quantifiable measures of the socioeconomic landscape and competitive environment surrounding each fandom.
Specifically, PCI, U, W, and GEO capture fundamental socioeconomic conditions of the team's supporter base at the city level, while MV is a quantitative representation of pre-season performance expectations.
This multidimensional approach allows us to explore how emotional dynamics interact with broader socioeconomic and competitive contexts.

Teams' market values (MV) are reasonable proxies for fandoms' expectations because they consider team's investments in key players, financial successes, brand strength, and overall reputation. This choice should guarantee a more reliable representation of fandoms' expectations.
Moreover, to evaluate the alignment between pre-season expectations and end-of-season outcomes, we consider the difference $\Delta_{mv}$ between a team's expected rank, based on its market value, $R_{mv}$, and its final rank, $R$, as $\Delta_{mv} = R_{mv} - R$.
If $\Delta_{mv}$ is negative, the team underperformed relative to its expected rank $R_{mv}$; otherwise, the team outperformed relative to it.
For example, since \textit{Spezia}, ranked first in terms of market value in Serie B, finished the season in 15th place, $\Delta_{mv}$ is -15, indicating that expectations were not met.
Conversely, since \textit{Catanzaro}, ranked 12th in market value, finished 5th, $\Delta_R$ is +7, reflecting a performance exceeding expectations.

Similar logic can be applied to further contextualize socioeconomic features.
For instance, wealth disparities can shape infrastructures, youth development, and team investments.
Thus, to evaluate an alignment between a rank based on such wealth factors and end-of-season outcomes, we consider the difference $\Delta_{pci}$ between a team's expected rank, in terms of per capita income,  $R_{PCI}$, and its final rank, $R$, as $\Delta_{pci} = R_{PCI} - R$.

\noindent Formally, each time series $P$ can be further described as $P=(\textbf{p},\textbf{a})$, where $\textbf{p}$ represents the emotional signals as sequence of observations, and \textbf{a} is a vector of $k$ attributes associated with the entire time series, i.e, contextual information that does not change over time.

\subsection{Time Series Clustering}
\label{subsec:clustering_method}

We aim to partition the time series dataset $\mathcal{P}$ defined in Section \ref{subsec:data} into clusters of emotionally aligned fandoms.
The dataset previously defined is multivariate, since each post contains the four emotional dimensions.
Without losing information, for this task we can work on the univariate time series, with $|\mathcal{E}|=1$, choosing \textit{joy}, being the most relevant dimension.

\begin{definition}[\textbf{Emotional Time Series Dataset Clustering}]
Clustering an emotional time series dataset $\mathcal{P}$ involves partitioning $\mathcal{P}$ into $k$ clusters, $\mathcal{C} = \{\mathcal{C}_1, \mathcal{C}_2, \dots, \mathcal{C}_k\}$, such that $\bigcup_{j=1}^k \mathcal{C}_j = \mathcal{P}$, i.e., all time series are assigned to a cluster, and $\mathcal{C}_i \cap \mathcal{C}_j = \emptyset \quad \text{for } i \neq j $, i.e., all clusters are disjoint.
\end{definition}

\noindent There exist several clustering methods for time series \cite{shumway2000time}, including adaptations of classic clustering algorithms, such as k-means and hierarchical clustering.
An important difference when dealing with time series is the distance calculation.
Euclidean distance is often insufficient because it can not handle temporal distortions.
A more relevant similarity measure is Dynamic Time Warping (DTW), which aligns time series by allowing flexible warping of the time axis.

In the experimental section, we use hierarchical clustering with DTW as the distance measure to partition the dataset.
Hierarchical clustering involves iteratively merging clusters based on pairwise DTW distances, producing a dendrogram for flexible analysis, cf. \ref{secA1.2}.
We also consider the k-means algorithm for time series, which partitions data by minimizing the within-cluster DTW distance to a representative centroid, updating cluster assignments iteratively. The full details and results of k-means clustering are reported in \ref{secA1.2} to avoid overloading the main experimental section.

Nothe that for this task, we further aggregate values per day, ensuring that all elements are conform to a daily temporal resolution, as teams have varying posting frequencies --- e.g., some teams may post 10 times a day while others only once.

\subsection{Burstiness and Memory}

\noindent Let $\tau$ be the inter-event time between two consecutive events. 
We aim to quantify whether the distribution of the inter-event times $P(\tau)$ deviate from the random activity pattern characterized by the exponential distribution.
In our cases, $P(\tau)$ is the distribution of the waiting time $\tau$ that elapses before a target emotion appears again in the emotional sequence.
We use the mean value of the distribution, $\mu_{\tau}=\frac{1}{n_\tau}\sum_{i=1}^{n_\tau}{\tau_i}$, where $n_\tau$ is the length of the sequence, and the standard deviation, $\sigma_{\tau}=\sqrt{\frac{\sum_{i=1}^{n_\tau}(\tau_i - \mu_\tau)^2}{n_\tau}}$, to measure the coefficient of variation $r$ as follows:

\begin{equation}
\centering
r=\frac{\sigma_{\tau}}{\mu_{\tau}}
\end{equation}

\noindent The variation $r$ is a measure of the dispersion of a distribution compared to its mean, useful to introduce the burstiness parameter $B$ as follows:

\begin{equation}
\centering
B=\frac{\sigma_{\tau}-\mu_{\tau}}{\sigma_{\tau}+\mu_{\tau}} = \frac{r-1}{r+1}
\end{equation}

\noindent Burstiness $B$ ranges from -1 and 1, where 1 indicates the bursty time series ($\sigma_{\tau}$ much higher than $\mu_{\tau}$), -1 the periodic one, and 0 indicates the random activity pattern.

Being $B$ affected by the finite number of events in the time series \cite{kim2016measuring}, it can be used a variation of $B$ for finite event sequences of size $n$ as follows:

\begin{equation}
\centering
B_n = \frac{{\sqrt{n+1} \cdot r - \sqrt{n-1}}}{{(\sqrt{n+1}-2)r + \sqrt{n-1}}}
\label{eq:burstiness}
\end{equation}

\noindent This modification allows to analyze time series with a relatively small number of events, which can be the case of low posting activity of a team account.
In the experiments, we will use the quantity $B_n$ described in Eq. \ref{eq:burstiness} to measure burstiness.

Moreover, we use the memory parameter $M$ \cite{goh2008burstiness} as follows:

\begin{equation}
\centering
M=\frac{1}{n-1}\sum_{i=1}^{n-1}\frac{(\tau_i-\mu_1)(\tau_{i+1}-\mu_2)}{\sigma_1\sigma_2},
\label{eq:memory}
\end{equation}

where $\mu_1$ and $\mu_2$, and $\sigma_1$ and $\sigma_2$ are sample mean and sample standard deviation of $\tau_i$'s values and $\tau_{i+1}$'s values, with $(i=1,...,n_{\tau}-1)$.
The memory coefficient $M$ is similar to the autocorrelation of a time series at lag=1.
Th measure can be extended to a generic lag $k$ \cite{schleiss2016two}, by replacing 1 in $n_{\tau}-1$ with $k$, and letting calculate the correlation between two inter-event times $\tau$ and $\tau'$ separated by $k$ events (instead of 1 event) to better capture long-range correlations.
Memory $M$ ranges from -1 and 1, where 1 indicates short/long inter-event times followed by short/long ones, -1 indicates short/long inter-event times followed by long/short ones, and 0 indicates no correlation.
In the experiments, we will use the quantity $M$ described in Eq. \ref{eq:memory} to measure the memory parameter, which is consistent with its original definition \cite{goh2008burstiness}.

\subsection{Regression}

To investigate the eventual relationship between emotional dynamics and external features, we employ multiple regression analysis.
Multiple regression is a statistical technique that is used to model the relationship between a dependent variable and multiple independent variables.
It extends simple linear regression by allowing the inclusion of several predictors, enabling the analysis of how each independent variable contributes to explaining the variation in the dependent variable while controlling for the effects of other variables.

The general form of a multiple regression model is:

\begin{equation}
Y = \beta_0 + \beta_1 X_1 + \beta_2 X_2 + \dots + \beta_p X_p + \epsilon,
\end{equation}

where $Y$ is the dependent variable, e.g., the final ranking $R$, and $X_p$ are the independent variables, such as the socioeconomic variables or the burstiness parameter as described in Eq. \ref{eq:burstiness}; $\beta_0$ is the intercept, $\beta_p$ are the coefficients representing the effect of each independent variable on $Y$, and $\epsilon$ is the error term, capturing unexplained variability. 

To evaluate the performance of the regression model, we use the coefficient of determination, $R^2$, and the root mean squared error, $RMSE$.

$R^2$ measures the proportion of variance in the dependent variable that is explained by the independent variables.
It ranges from 0 to 1, where higher values indicate better fit.
$R^2$ is calculated as:

\begin{equation}
    R^2 = 1 - \frac{\text{SS}_{\text{res}}}{\text{SS}_{\text{tot}}},
\end{equation}

where $\text{SS}_{\text{res}}$ is the residual sum of squares and $ \text{SS}_{\text{tot}} $ is the total sum of squares.

$RMSE$ measures the average deviation of the predicted values from the actual values.
It provides a sense of the model's prediction error in the units of the dependent variable.
$RMSE$ is calculated as:

\begin{equation}
    \text{RMSE} = \sqrt{\frac{1}{n} \sum_{i=1}^n (Y_i - \hat{Y}_i)^2},
\end{equation}

where $Y_i$ is the actual value, $\hat{Y}_i$ is the predicted value, and $n$ is the number of observations. Lower $RMSE$ values indicate better predictive accuracy.

\begin{figure}[t!]
\centering
\includegraphics[scale=0.5]{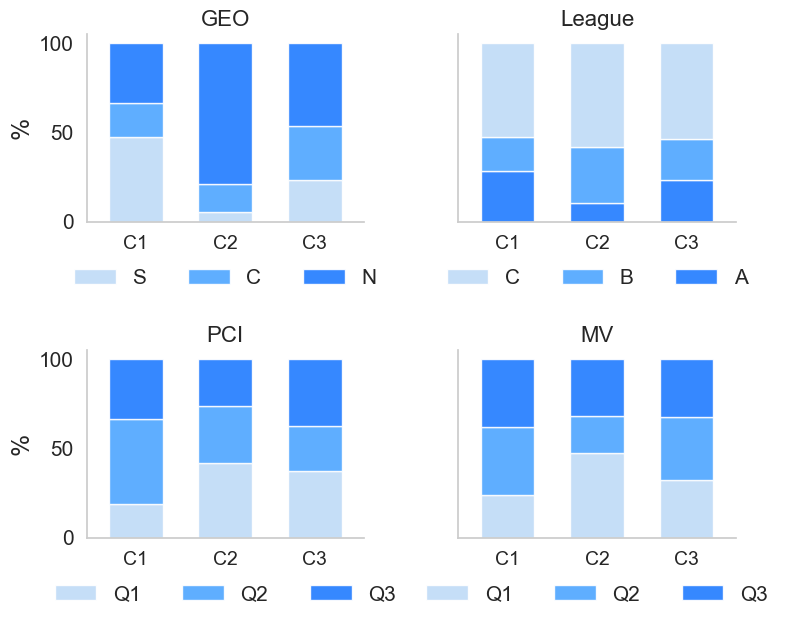}
\caption{Distribution of categorical features across the three clusters identified by the hierarchical clustering. The labels \textit{S, C, N} in the GEO legend represent \textit{South, Center, and North Italy}, while \textit{C, B, A} in the League legend stands for Lega Pro, Serie B, and Serie A,  and \textit{Q1, Q2, Q3} correspond to the tertiles of the binned average per capita incomes (PCI) and market values (MV), with \textit{Q1} representing the lowest category and \textit{Q3} the highest.}
\label{fig:clusters_cat}
\end{figure}

\begin{figure}[t!]
\centering
\includegraphics[scale=0.20]{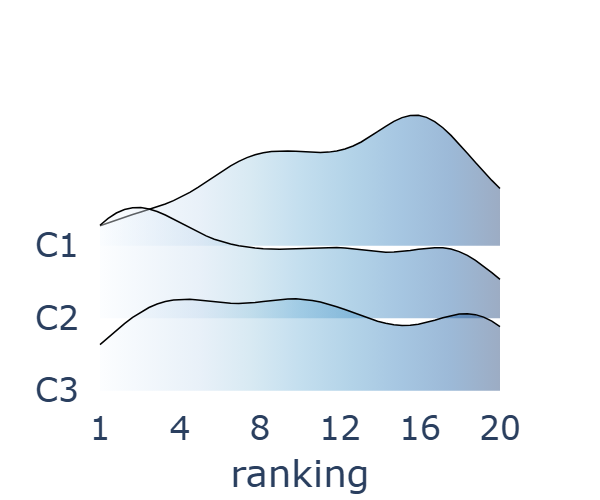}
\includegraphics[scale=0.20]{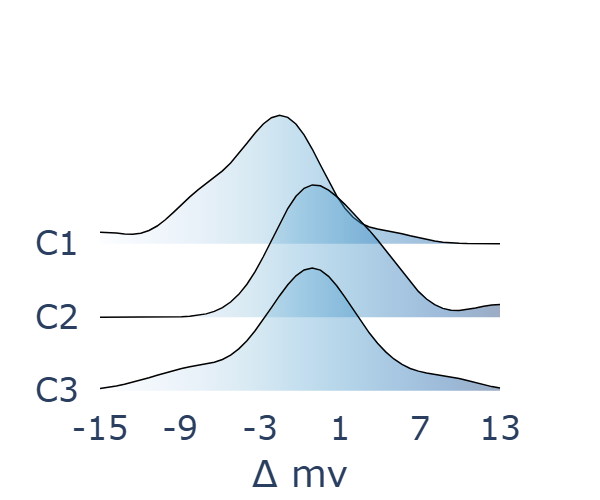}
\includegraphics[scale=0.20]{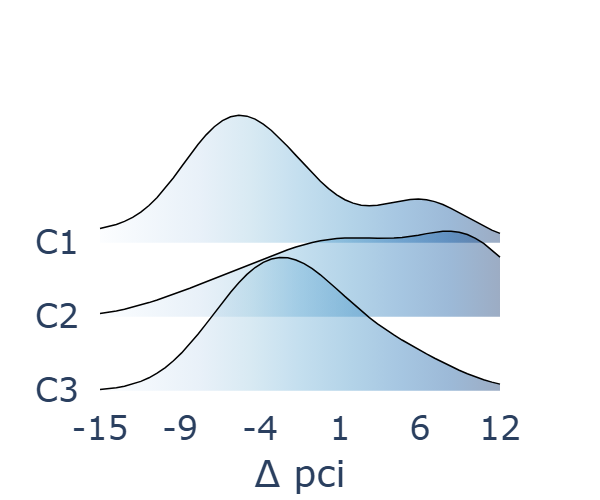}
\caption{Distribution of continuous features across the three clusters identified by the hierarchical clustering.}
\label{fig:clusters_cont}
\end{figure}

\section{Results}
\label{sec:results}

\subsection{Common Features within Clusters}
\label{subsec:clust_res}

While partitioning the dataset, we expect that fandoms within the same cluster exhibit similar trends in their emotional behavior compared to fandoms in other clusters.
As mentioned in Section \ref{subsec:clustering_method}, we apply the hierarchical clustering algorithm, selecting $k=3$ number of clusters by looking at the resulting dendrogram, cf. \ref{secA1.3}, while the results for the k-means clustering are left for \ref{secA1.3}.

Figure \ref{fig:clusters_cat} and Figure \ref{fig:clusters_cont} highlight the feature distributions in the three identified clusters.
Indeed, to gain insights into such groups, we analyze how features, as described in Section \ref{subsec:features}, distribute within groups, recalling that clustering does not ensure or aims to maximize feature homogeneity, as it operates exclusively on the emotional time series data.
First of all, from Figure \ref{fig:clusters_cat}, there emerges a highly homogeneous cluster in terms of geographical distribution --- cluster id, \textit{C2} --- which consists of 15 fandoms from North Italy, and only 3 and 1 fandoms from Center and South Italy, respectively; these are \textit{Empoli}, \textit{Perugia}, \textit{Pontedera}, from Center, and \textit{Catanzaro}, from South.
Interestingly, only 2 fandoms from Serie A are present in this cluster: \textit{Empoli} and \textit{Monza}.
Among the 6 teams from Serie B, five of them --- \textit{Parma, Como, Venezia, Cremonese, and Catanzaro} --- are ranked in the top six of the league's final ranking, plus \textit{Cittadella}.
The socioeconomic attributes of this cluster falls, on average, within the lowest tertile, Q1, with the same applying to MV, also in Q1.
Moreover, teams within this group, on average, have finished in the top part of the ranking, and generally show higher $\Delta_{pci}$, and, to a lesser extent, higher $\Delta_{mv}$ (Figure \ref{fig:clusters_cont}).
Remember that positive differences in both variables indicate final results that exceeded pre-season expectations.

Moving on to the other groups, the cluster labeled with id \textit{C1} consists of 7, 4 and 10 fandoms from North, Center and South Italy, respectively, whereas cluster id \textit{C3} consists of 20, 13 and 10 fandoms from North, Center, and South Italy, respectively (Figure \ref{fig:clusters_cat}).
In other words, the geographical distribution reveals a higher concentration of teams from the South in \textit{C1}, and from the North in \textit{C3}.
There is no notable homogeneity in terms of leagues, suggesting that emotional differences are not influenced by league hierarchies.
Regarding PCI and MV, the two clusters show mixed distributions without clear patterns, though the first one contains a higher proportion of Q2 bins.
As seen in Figure \ref{fig:clusters_cont}, it is evident that the first cluster, \textit{C1}, predominantly features negative moods, with both the values of $\Delta_{pci}$ and $\Delta_{mv}$ falling below zero.
These negative differences indicate that final results did not meet pre-season expectations. 
The third cluster, \textit{C3}, displays less homogeneity, except for a slight negative mood in $\Delta_{pci}$.

\begin{figure}[t!]
\centering
\includegraphics[scale=0.48]{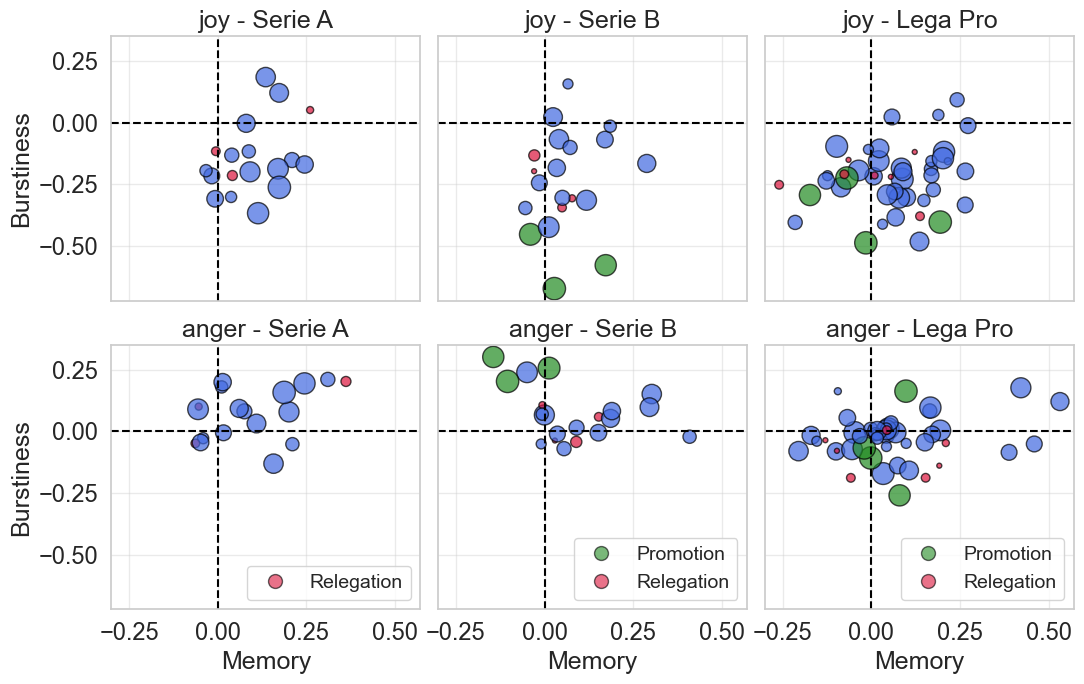}
\caption{Focus on \textit{joy} and \textit{anger}: Points represent pairs of burstiness parameter and memory coefficient values. Sizes map the final rank of the teams; moreover, green/red points highlight a promotion/relegation to a higher/lower league.}
\label{fig:scatters}
\end{figure}

\subsection{Loyalty as Anti-Bursty Patterns and Memory Effects}
\label{subsec:burst_res}

In this section, we investigate whether fandoms' emotional signals manifest bursty properties as other human dynamics \cite{karsai2018bursty}, or whether they exhibit random, or non-random but regular, dynamics.
Figure \ref{fig:scatters} shows burstiness vs. memory diagrams, grouped by leagues, for two target emotions: joy and anger.
We do not consider sadness and fear, in this main analysis, since they rarely come as the most frequent emotions within posts, see \ref{secA1.2}.

Joy and anger unveil different event-based dynamics. 
Joy rarely displays a bursty trend, and only a few fandoms have a burstiness greater than 0.
In Serie A, these teams are \textit{Fiorentina}, \textit{Lazio}, and \textit{Salernitana}, the latter one also relegated to Serie B at the end of the season.
In Serie B, the teams are \textit{Bari} and \textit{Sampdoria}.
In Lega Pro, \textit{Ancora}, \textit{Catania}, \textit{Foggia}, and \textit{SPAL}.
Notably, these fandoms also manifest a positive memory effect.
The other fandoms, in all leagues, overall depict anti-bursty emotional dynamics with respect to joy.
This can indicate consistent loyalty and supportive behavior for the respective teams.
Serie B has more supportive fandoms, also including the four teams relegated to Lega Pro at the end of the season.
This suggests a complex bond with the teams that goes beyond the final rankings and match results, possibly linked to pre-season expectations, cf. \ref{subsec:clust_res}.

In contrast to joy, anger displays more heterogeneous behaviors, with memory effects varying more widely across teams.
Only a few fandoms highlight a burstiness value less than 0 among Serie A and Serie B, namely \textit{Lazio}, and \textit{Udinese} in Serie A, and \textit{Bari}, \textit{Cittadella}, \textit{Lecco}, \textit{Pisa}, and \textit{Ternana} in Serie B, with \textit{Lecco} and \textit{Ternana} being relegated to Lega Pro, and \textit{Bari} having disputed the play-out matches against \textit{Ternana}.
These behaviors are consistent with the bursty patterns observed for joy for these teams, e.g., \textit{Lazio} has both the highest values of joy-burstiness and the lowest values of anger-burstiness.
This indicates a constant feeling of anger, while joy can be relative to single matches.
Notably, Lega Pro's burstiness values differ from the other two leagues.
An anti-bursty anger behavior is observed among the relegated teams.
Moreover, memory with respect to anger uncovers a wide heterogeneous behavior, making it difficult to identify a common pattern across all fandoms.
In Serie B, for instance, \textit{Como}, \textit{Cremonese} and \textit{Venezia} show the most negative memory effect, whereas \textit{Cittadella}, \textit{Palermo} and \textit{Sampdoria}, the most positive memory effect.

\subsection{Correlating Emotional Events and Fans' Expectations to Final Rankings}
\label{subsec:regress_ress}

\begin{figure}[t!]
\centering
\includegraphics[scale=0.69]{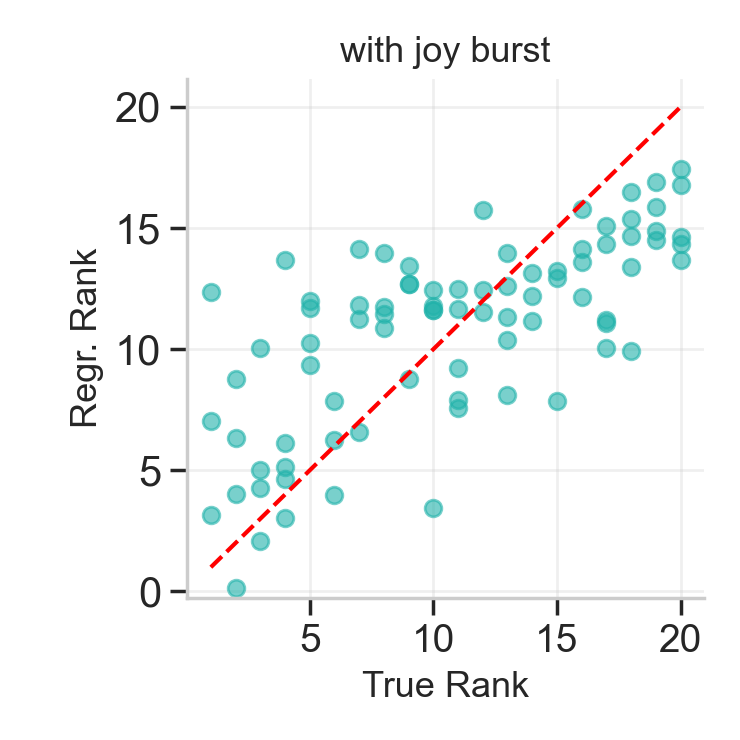}
\includegraphics[scale=0.69]{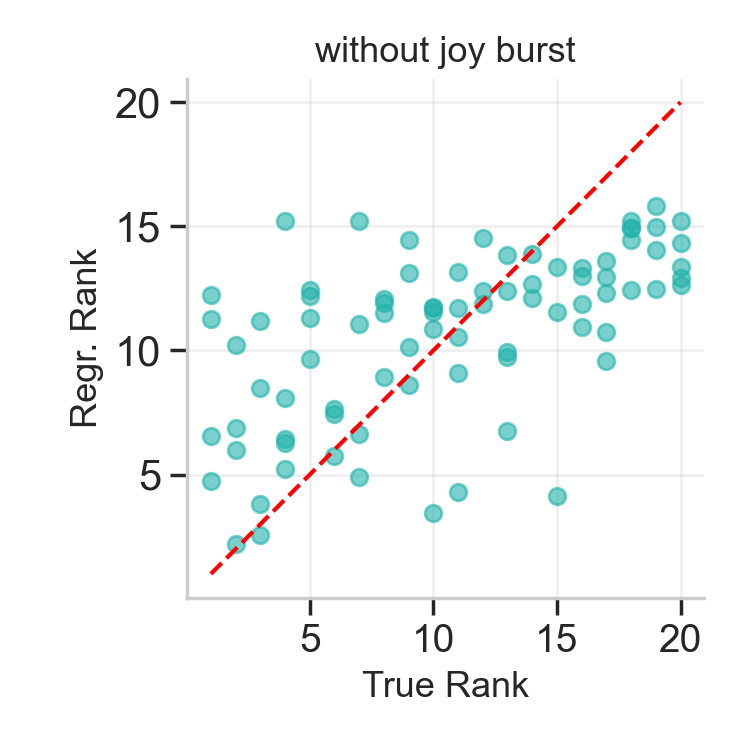}
\caption{Multi-regression with and without using burstiness in joy as a dependent variable in predicting final rankings.}
\label{fig:regr}
\end{figure}

In this section, we correlate, with multi-regression, the following dependent variables with \textit{rankings}, this latter one used as the independent variable: joy burstiness $B_{joy}$, team heritage $R_H$, $PCI$, $MV$, $W$. Formally, our full model has the form

\begin{equation}
    R \sim  \beta_0 + \beta_1 R_{H} + \beta_2 PCI + \beta_3 MV + \beta_4 W + \beta_5 B_{joy} + \epsilon
\end{equation}

We began with all variables and systematically removed those that did not significantly affect the model's performance, as the overall results remained consistent without them.
Doing this, we observed that removing joy burstiness from the model led to a decrease in performance.
Formally, this happens when the model has the form

\begin{equation}
    R \sim  \beta_0 + \beta_1 R_{H} + \beta_2 PCI + \beta_3 MV + \beta_4 W + \epsilon
\end{equation}

Figure \ref{fig:regr} sums up the results, demonstrating how much burstiness is a highly relevant correlate of team performance.
The model excluding burstiness in joy (the right one in the figure) shows a decrease of $32\%$ in $R^2$, and $15.57\%$ in $RMSE$, being achieved $R^2=0.50$ and $RMSE=4.11$ with the model including burstiness (the left one in the figure), and $R^2=0.34$ and $RMSE=4.75$ with the one excluding it. 
When joy burstiness is included as a predictor, the model achieves better performance.

\section{Discussion and Conclusion}
\label{sec:disc}

We analyzed the complex emotional dynamics of online soccer fandoms in Italy's Serie A, B, and Lega Pro, processing comments from Instagram to quantify users' emotions, such as joy and anger, throughout the 2023-24 season.
Using clustering for partitioning the resulting emotional time series, we uncovered groups of fandoms featuring similar pre-season expectations and geographical distribution, exclusively by looking at such emotional signals.
Then, we used measures from complex systems analysis, namely burstiness and memory \cite{goh2008burstiness}, unveiling, overall, bursts of anger and steady feelings of joy across the 83 fandoms considered.
Finally, we demonstrated that burstiness in joy is a relevant factor when correlating teams' final ranking, in multi-regression where also pre-season expectations and socioeconomic features have been considered.

Although correlates such as goals scored, expected goals (xG)\footnote{\url{https://understat.com/}}, and complex statistics that capture offensive and defensive effectiveness \cite{cintia2015harsh} are already established predictors of team rankings, our research focused on the other side of the equation: patterns of fan behavior.
Rather than attempting to replace previous predictors, we investigated whether social characteristics of fan communities could also correlate with team outcomes, especially given that both the clustering and burstiness analyses revealed non-random and meaningful emotional patterns.
The current analysis is confined to a single season, and one year alone might provide a restricted view of the potential long-term burstiness in users' behavior.
However, including team heritage as a dependent variable does not improve the correlations, reaching $R^2=0.47$ when removing it, compared to $R^2=0.50$ as outlined in Section \ref{subsec:regress_ress}.
Hence, in the future, we plan to incorporate additional features that may more accurately reflect team traditions and other complex fan characteristics, e.g., more refined metrics that can account for fan pre-season expectations \cite{wang2023making}.
Moreover, the bursty nature of emotional dynamics has received limited attention outside cognitive and psychological studies, with only preliminary investigations previously conducted \cite{citraro2024burstiness}.
For future research, we also plan to expand the investigation of emotional burstiness by integrating psychological frameworks as well as exploring the neural correlates of fan engagement in sports \cite{duarte2017tribal}.
An even more interdisciplinary approach will provide deeper insights into how collective emotions manifest in sports communities and potentially influence team performance outcomes.

The quality of our results inevitably depends on the emotion classifier used to label comments, which is based on the \textit{feel-it} corpus for Italian language \cite{bianchi2021feel}.
Since this corpus is derived from social media data, it aligns well with the content of our dataset.
However, the limitation to only four emotions may narrow the spectrum of possible engagement experienced in sport contexts.
Although we mainly observed collective expressions of joy and anger, these could still be interpreted in different ways.
For example, joy may arise not only from one's own team winning but also from a rival team losing.
In the future, we plan to further explore these emotional shades, although doing so requires a more complex understanding of language.

Finally, while burstiness in joy strongly correlates with final rankings, other variables --- such as socioeconomic wealth (W), market value (MV), and even fans' pre-season expectations (PCI) --- show weaker effects.
This suggests that burstiness may capture the changing and dynamic \textit{flow} of fan engagement, one that static features cannot fully reflect.
PCI and W represent more structural or contextual factors that may lack the real-time granularity needed to reflect emotional momentum or shifts in collective sentiment throughout the season.

To conclude, although online fandom is by no means comparable to the physical experience inside and outside stadiums, the digital traces left by fans on social media suggest potential avenues for research in sports analytics, social dynamics, and fan engagement studies.

\section*{List of Abbreviations}

\begin{itemize}
\item $B$: Burstiness parameter
\item $DTW$: Dynamic Time Warping
\item $GEO$: Geographical location
\item $M$: Memory parameter
\item $MV$: Market value
\item $PCI$: Per Capita Income (City)
\item $R$: Final rank
\item $R_H$: Team heritage
\item $R_{mv}$: Expected rank based on market value
\item $R_{PCI}$: Expected rank based on per capita income
\item $RMSE$: Root Mean Squared Error
\item $U$: Unemployment rate
\item $W$: Welfare spending
\item $xG$: Expected goals
\item $\Delta_{mv}$: Difference between expected rank (market value) and final rank
\item $\Delta_{pci}$: Difference between expected rank (PCI) and final rank
\end{itemize}

\section*{Availability of Data and Materials}

Data used in this work are available at the following link: \url{https://zenodo.org/records/15350266}

\section*{Competing interests}
The authors declare that they have no competing interests.

\section*{Author Contributions}

Conceptualization: SC, GM; Data curation: SC; Formal analysis: SC, GM; Investigation: SC, GM, EF; Methodology: SC, GM; Supervision: EF; Validation: SC; GM, EF; Visualization: SC, GM; Writing, original draft: SC, GM, EF.

\section*{Acknowledgments}

This work is supported by SoBigData.it which receives funding from the European Union – NextGenerationEU – National Recovery and Resilience Plan (Piano Nazionale di Ripresa e Resilienza, PNRR) – Project: “SoBigData.it – Strengthening the Italian RI for Social Mining and Big Data Analytics” – Prot. IR0000013 – Avviso n. 3264 del 28/12/2021.  
\\ \ \\
\noindent We dedicate this work to Giorgio Vignando, a quiet sentinel of the midfield. He taught us that football’s joy does not simply live in the roar of goals, but in the unspoken beauty of a battle fought in shadow.

\bibliography{bibliography}

\newpage


\begin{appendices}

\section{Notes on Materials and Methods}\label{secA1}

\begin{figure}[t!]
\centering
\subfloat[]{\includegraphics[scale=0.28]{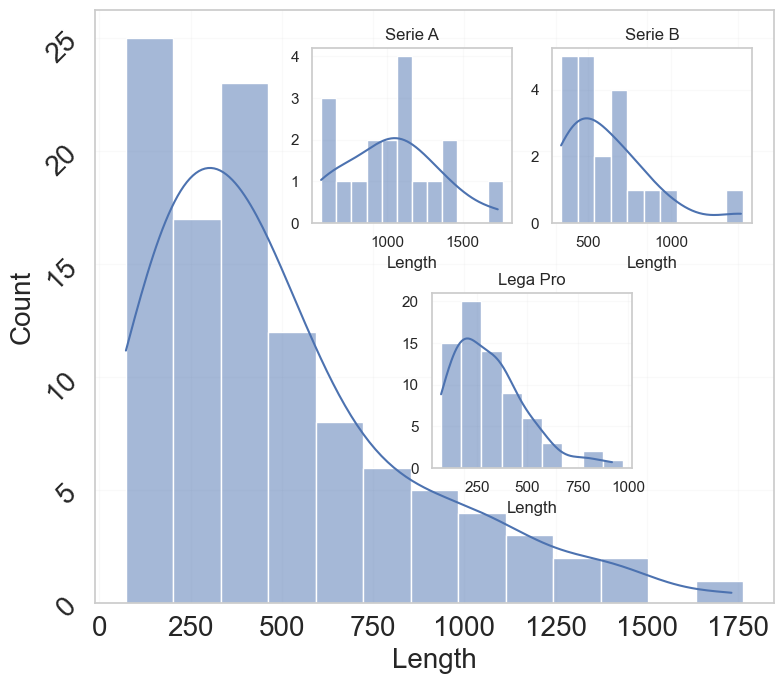}}
\subfloat[]{\includegraphics[scale=0.28]{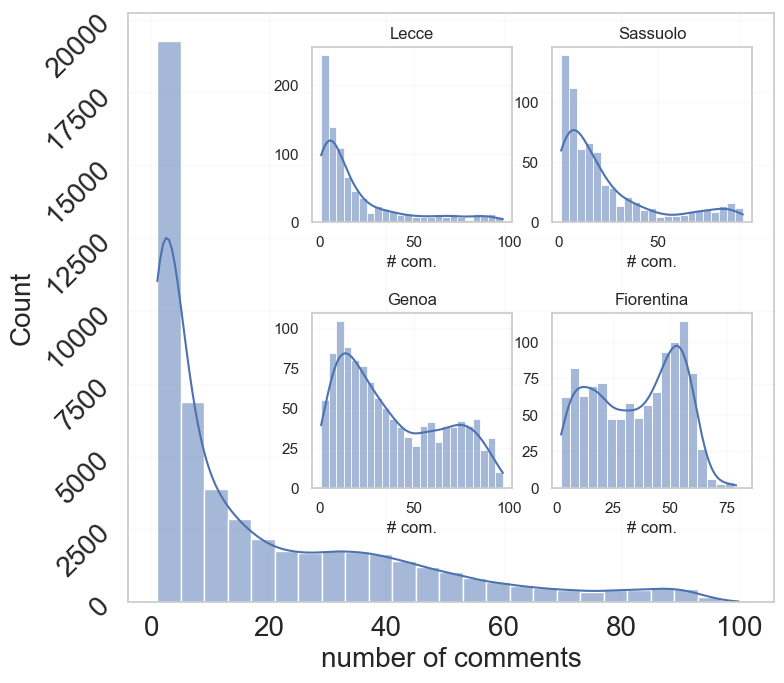}}
\qquad
\subfloat[]{\includegraphics[scale=0.28]{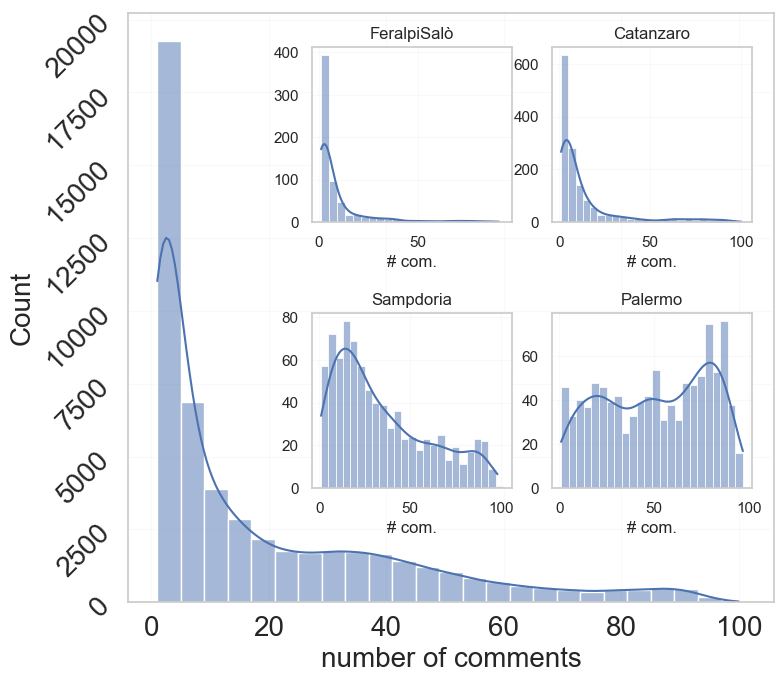}}
\subfloat[]{\includegraphics[scale=0.28]{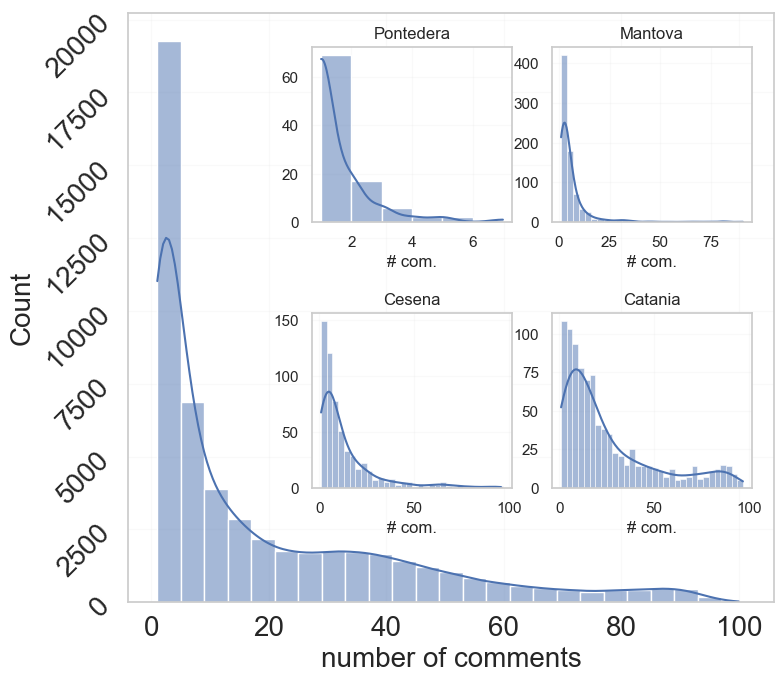}}
\caption{(a) Distribution of time series lengths, where the length is given by the number of teams' posts over the 2023/24 seasons. The inner plots show the distributions of each league; (b-d) Distribution of user comments within posts, where the inner plots show the individual distributions of randomly selected teams across all leagues to show the strong heterogeneity between them.}
\label{fig:post_length}
\end{figure}

\subsection{Posts and Comments}
\label{secA1.1}

The frequency of posting activity varies significantly between teams.
For instance, one team might post multiple times a day, while another might post only once or twice a week.
This disparity is illustrated in Fig. \ref{fig:post_length} (a).
The inner plots also indicate that, on average, teams from Serie A tend to post more frequently than teams from other leagues.
In any case, in our analyses, we keep these teams' lengths when calculating the burstiness and memory coefficients.
Instead, another pre-process step is needed to reduce noise when performing clustering, that is the temporal alignment of time series.
This involves aggregating posts on a day-by-day basis from the start to the end of the considered seasons.
Hence, when computing the percentage of emotions, we further average these percentages on the daily basis.

Another relevant distribution to consider is the number of user comments on posts.
This distribution is also long-tailed, as shown in Fig. \ref{fig:post_length} (b-d). 
Since a post can have only one or just a few comments, it is necessary to reduce this noise.
In other words, detecting 100\% joy in a post should not be meaningful if the post contains only a single comment.
However, as indicated by the inner plots, the overall distribution does not accurately represent the individual distributions.
For example, the fan bases of Catania and Palermo engage more than those of FeralpiSalò and Pontedera, because their user pools differ a lot. 
Therefore, to account for this heterogeneity, we apply individual thresholds to select the posts that are retained: we consider only those posts that have a number of comments equal to or greater than the median of the distribution for each individual account.

\begin{figure}[t!]
\centering
\subfloat[]{\includegraphics[scale=0.38]{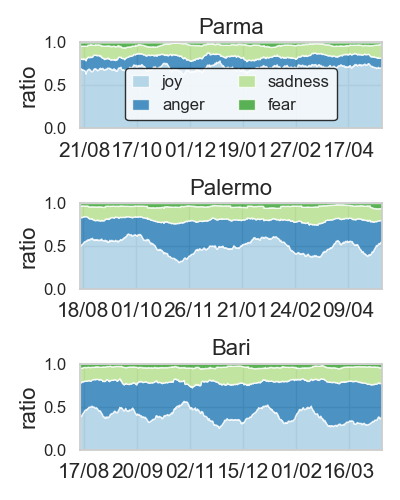}}
\subfloat[]{\includegraphics[scale=0.38]{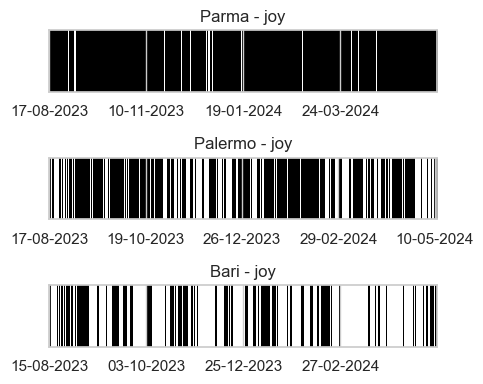}}
\subfloat[]{\includegraphics[scale=0.38]{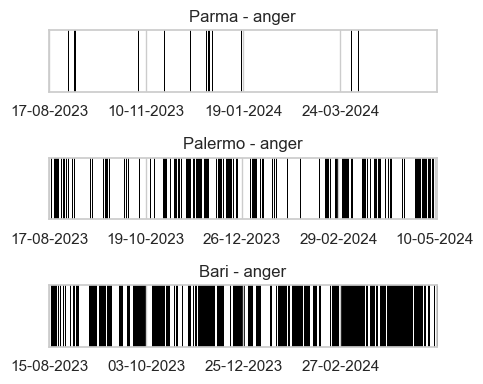}}
\caption{From real-value based time series (a) to event-based time series according to different emotions, i.e., joy (b) and anger (c).}
\label{fig:example_reprs}
\end{figure}

\begin{figure}[t!]
\centering
\includegraphics[scale=0.75]{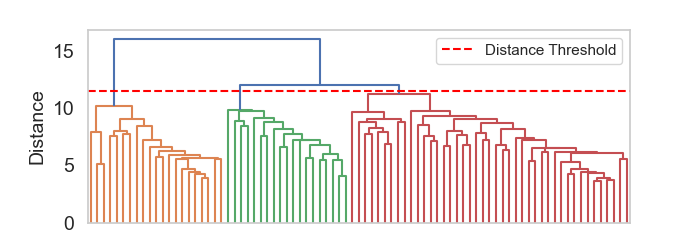}
\caption{The dendrogram obtained by performing hierarchical clustering on the real-value based time series dataset. The red line indicates the distance at which we cut to obtain $k=3$ clusters.}
\label{fig:dendr}
\end{figure}

\subsection{Event-based Time Series}
\label{secA1.2}

Fig. \ref{fig:example_reprs} highlights the event-based time series representation.
We consider three teams from Serie B as a toy example, but the following holds true on average for all the fandoms in the dataset.
Notably, the two dominant emotions expressed throughout the season are joy and anger, which is why we focus exclusively on these two emotions.
The resulting event-based time series emphasize the occurrences of joy and anger events.
Parma, for instance, consistently at the top of the rankings, exhibits a joy-based time series dominated by events, and the anger-based one with occasional bursts events.
Palermo shows very important inhomogeneous patterns for both emotions, while Bari is consistently inclined towards anger-based events.

\begin{figure}[t!]
\centering
\includegraphics[scale=0.5]{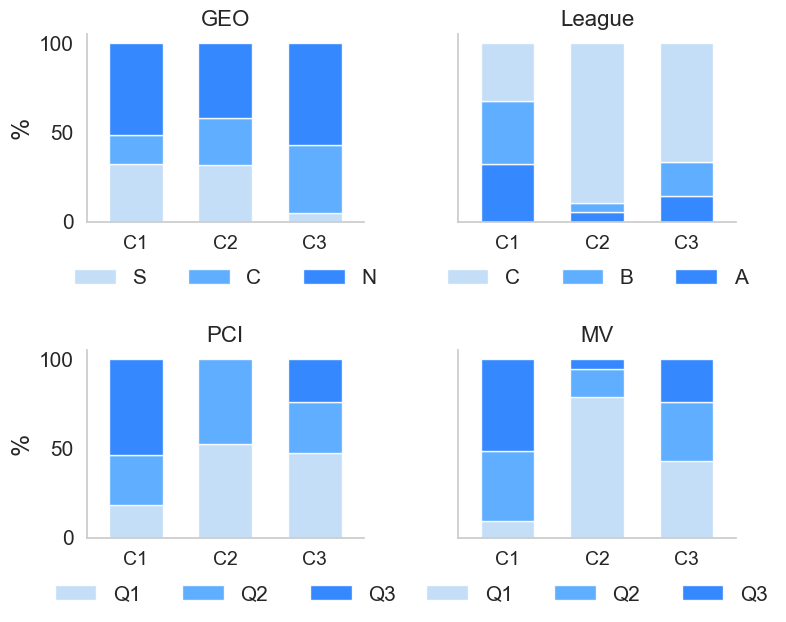}
\caption{Distribution of categorical features across the three clusters identified by the k-means clustering. The labels \textit{S, C, N} in the GEO legend represent \textit{South, Center, and North Italy}, while \textit{C, B, A} in the League legend stands for Lega Pro, Serie B, and Serie A,  and \textit{Q1, Q2, Q3} correspond to the tertiles of the binned average per capita incomes (PCI) and market values (MV), with \textit{Q1} representing the lowest category and \textit{Q3} the highest.}
\label{fig:clusters_cat_kmeans}
\end{figure}

\begin{figure}[t!]
\centering
\includegraphics[scale=0.20]{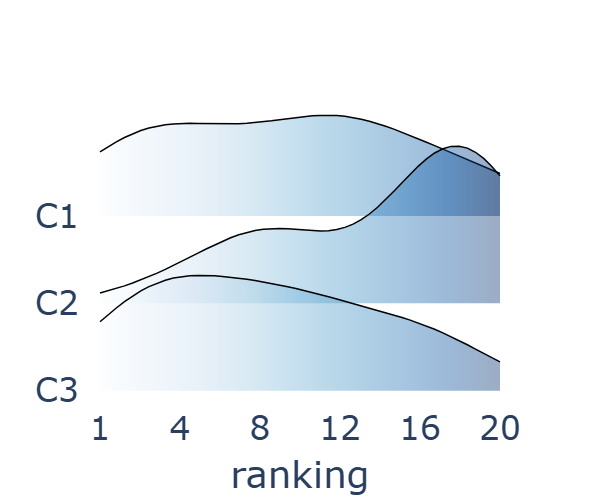}
\includegraphics[scale=0.20]{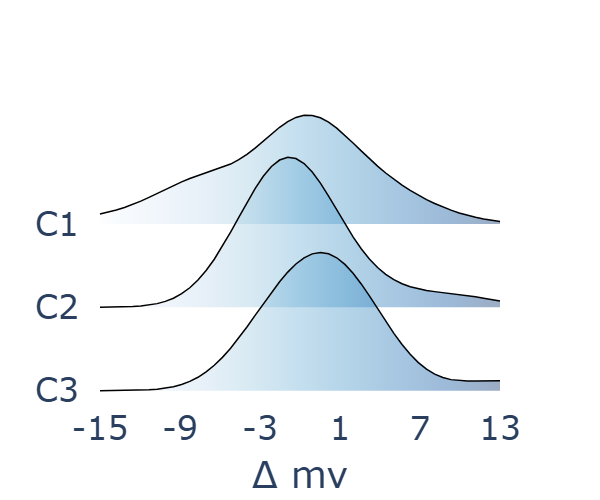}
\includegraphics[scale=0.20]{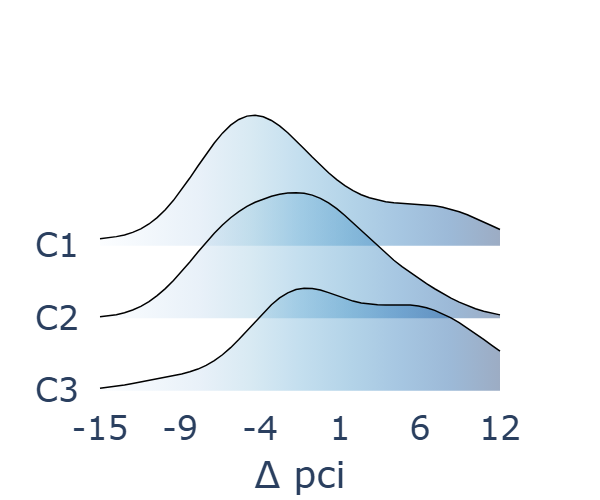}
\caption{Distribution of continuous features across the three clusters identified by the k-means clustering.}
\label{fig:clusters_cont_kmeans}
\end{figure}

\subsection{Notes on Clustering}
\label{secA1.3}

Figure \ref{fig:dendr} shows the dendrogram resulting from hierarchical clustering.
The plot clearly indicates a division into $k=3$ clusters, with the optimal cut made at a distance of $d=10.5$.

This section also presents the results of k-means clustering, where we again chose $k=3$ as the optimal number of clusters.
As shown in Figures \ref{fig:clusters_cat_kmeans} and \ref{fig:clusters_cont_kmeans}, there are notable differences compared to the hierarchical clustering results described in the main article.
However, the differences do not indicate contradictory analyses, but rather highlight the ability of the two data partitioning methods to identify different patterns.
With kmeans, indeed, a highly homogeneous cluster emerges in terms of the league --- cluster id, \textit{C2}.
It contains only one team fandom from Serie A, \textit{Empoli}, and one from Serie B, \textit{FeralpiSalò}, out of 19 fandoms.
Interestingly, most of the fans within this cluster fall into the lowest categories for both PCI and MV.
For instance, the three teams that do not fall into Q1 tertile acording to MV are \textit{Empoli}, \textit{FeralpiSalò}, and \textit{Crotone}.
The \textit{C2} cluster predominantly contains teams ranked in the lower part of the standings.

The two clustering methods provide complementary insights: while hierarchical clustering emphasizes geographic patterns, k-means tends to capture league-based distinctions more effectively.
Both methods succeed in identifying meaningful characteristics — notably, their ability to detect patterns in external features by analyzing similarities in emotional time series alone.

\end{appendices}

\end{document}